\documentclass{revtex4}
\usepackage{amssymb}
\usepackage{amsfonts}
\usepackage{amsmath}

\input{tcilatex}

\begin{document}

\title{Position-dependent mass charged particles in magnetic and
Aharonov-Bohm flux fields: separability, exact and conditionally exact
solvability}
\author{ Zeinab Algadhi}
\email{zeinab.algadhi@emu.edu.tr}
\author{Omar Mustafa}
\email{omar.mustafa@emu.edu.tr}
\affiliation{Department of Physics, Eastern Mediterranean University, G. Magusa, north
Cyprus, Mersin 10 - Turkey,\\
Tel.: +90 392 6301378; fax: +90 3692 365 1604.}

\begin{abstract}
\textbf{Abstract:} Using cylindrical coordinates, we consider
position-dependent mass (PDM) charged particles moving under the influence
of magnetic, Aharonov-Bohm flux, and a pseudoharmonic or a generalized
Killingbeck-type potential fields. We implement the PDM-minimal-coupling\
recipe \cite{26}, along with the PDM-momentum operator \cite{27}, and report
separability under radial cylindrical and azimuthal symmetrization settings.
For the radial Schr\"{o}dinger part, we transform it into a radial
one-dimensional Schr\"{o}dinger-type and use two PDM settings, $g\left( \rho
\right) =\eta \rho ^{2}$ and $g\left( \rho \right) =\eta /\rho ^{2}$, to
report on the \emph{exact solvability} of PDM charged particles moving in
three fields: magnetic, Aharonov-Bohm flux, and pseudoharmonic potential
fields. Next, we consider the radial Schr\"{o}dinger part as is and use the
biconfluent Heun differential forms for two PDM settings, $g\left( \rho
\right) =\lambda \rho $ and $g\left( \rho \right) =\lambda /\rho ^{2}$, to
report on the \emph{conditionally exact solvability} of our PDM charged
particles moving in three fields: magnetic, Aharonov-Bohm flux, and
generalized Killingbeck potential fields. Yet,\ we report the spectral
signatures of the one-dimensional $z$-dependent Schr\"{o}dinger part on the
overall eigenvalues and eigenfunctions, for all examples, using two $z$%
-dependent potential models (infinite potential well and Morse-type
potentials).

\textbf{PACS }numbers\textbf{: }03.65.-w, 03.65,Ge, 03.65.Fd

\textbf{Keywords:} Quantum position-dependent mass Hamiltonian, PDM-momentum
operator, cylindrical coordinates, magnetic and Aharonov-Bohm flux fields,
pseudoharmonic potential, Killingbeck potential. 
\end{abstract}

\maketitle

\section{Introduction}

In the non-relativistic Schr\"{o}dinger equation, a position-dependent
deformation of the mass of the particle, or a position-dependent deformation
of the coordinates involved, are two parallel assumptions that yield , in
short, to position-dependent mass (PDM) concept. This concept have attracted
much attention in the literature over the years, for both classical and
quantum mechanical systems (a sample of references can be found in e.g., 
\cite%
{1,2,3,4,5,6,7,8,9,10,11,12,13,14,15,16,17,18,19,20,21,22,23,24,25,26,27}).
Hereby, the most prominent PDM non-relativistic Hamiltonian is known as the
von Roos Hamiltonian \cite{1} (in $\hbar =2m_{\circ }=1$ units) 
\begin{equation}
\hat{H}=-\frac{1}{4}\left[ M\left( \overrightarrow{r}\right) ^{a}%
\overrightarrow{\nabla }M\left( \overrightarrow{r}\right) ^{b}.%
\overrightarrow{\nabla }M\left( \overrightarrow{r}\right) ^{c}+M\left( 
\overrightarrow{r}\right) ^{c}\overrightarrow{\nabla }M\left( 
\overrightarrow{r}\right) ^{b}.\overrightarrow{\nabla }M\left( 
\overrightarrow{r}\right) ^{a}\right] +V\left( \overrightarrow{r}\right) .
\end{equation}%
Where $M\left( \overrightarrow{r}\right) =m_{\circ }m\left( \overrightarrow{r%
}\right) $, $m_{\circ }$ is the rest mass, $m\left( \overrightarrow{r}%
\right) $ is a position-dependent dimensionless scalar multiplier that forms
the position-dependent mass $M\left( \overrightarrow{r}\right) $, $V\left( 
\overrightarrow{r}\right) $ is the potential field, and $a,b,c$ are called
the ambiguity parameters that satisfy von Roos constraint $a+b+c=-1$. Yet,
this Hamiltonian is known to be associated with an ordering ambiguity
problem as a result of the non-unique representation of the kinetic energy
operator. An obvious radical change in the profile of the kinetic energy
term occurs when the values of the ambiguity parameters are changed
(consequently, the profile of the effective potential will radically
change). There exist an infinite number of ambiguity parametric settings
that satisfy the von Roos constraint above. In the literature, however, one
may find many suggestions on the ambiguity parametric values \ \cite%
{2,3,4,5,6,7,8,9,10,11,12,13,14}. Yet, the only physically acceptable
condition (along with the von Roos constraint) on the ambiguity parameters
is that $a=c$ to ensures continuity at the abrupt heterojunction (e.g.,
Refs. \cite{15,16} ). The rest are based on different eligibility proposals
which are, at least, mathematically challenging and useful models that
enrich the class of exactly solvable or conditionally exactly solvable
quantum mechanical systems \cite{17,18,19,20,21,22,23,24,25,26}.
Nevertheless, it was only very recently that a PDM momentum operator is
constructed by Mustafa and Algadhi \cite{27} and resulted in fixing the
ordering ambiguity parameters at $a=c=-1/4$ and $b=-1/2$ (known in the
literature as Mustafa and Mazharimousavi's paramtric settings \cite{13}).

On the other hand, quantum mechanical charged particles of constant mass
moving in a uniform magnetic field (with some occasional inclusion of the
Aharonov-Bohm flux field) have been a subject of research interest over the
years (e.g., see the sample of references \cite{28,29,30,31,32,33,34} and
related references cited therein). Only a handful number of attempts were
made to treat PDM charged particles in uniform magnetic field \cite%
{24,25,26,35}. \ Nevertheless, to facilitate exact solvability for PDM
charged particles in electromagnetic fields, Mustafa and Algadhi \cite{27}
have used some canonical point transformation that maps the PDM Hamiltonian
into conventional constant mass setting. In so doing, the exact solutions
for PDM systems are inferred from those for conventional constant mass
systems. Moreover, in their very recent study, Eshghi et al. \cite{35} have
used Ben Danial and Duke's parametric settings $a=c=0$ and $b=-1$ (c.f.,
e.g., \cite{2,3,4,22,23,24}) and considered PDM-charged particles moving in
both magnetic and Aharonov-Bohm flux fields.

In the current methodical proposal, however, we use the freshly constructed
PDM momentum operator of Mustafa and Algadhi \cite{27} and study, within
cylindrical and azimuthal symmetrization settings, the PDM-charged particles
moving under the influence of magnetic and Aharonov-Bohm flux fields. The
PDM minimal coupling\ \cite{26} is used in the process. We also explore the
separability of the corresponding PDM-Schr\"{o}dinger equation, along with
its feasible exact and conditionally exact solvability through two sets of
illustrative examples. To the best of our knowledge, and within the current
methodical proposal setting, no such studies have been carried out and/or
are available in the literature so far.

The organization of this paper is in order. In section 2, we use the
PDM-minimal-coupling\ recipe \cite{26}, along with the PDM-momentum operator 
\cite{27}, and discuss the separability of the problem within the
cylindrical coordinates $\left( \rho ,\varphi ,z\right) $, indulging
azimuthal symmetry. A purely radial $\rho $-coordinate dependent, (12)
below, and a simplistic one-dimensional $z$-dependent, (11) below, Schr\"{o}%
dinger equations resulted in the process. In connection with the radial $%
\rho $-dependent part, moreover, we choose to treat it in two different
ways. The first of which is to transform it into a radial one-dimensional
Schr\"{o}dinger form and discuss, in section 3, its exact solvability using
a pseudoharmonic potential (which is usually used for quantum dots and
antidotes, e.g., \cite{31,32,33}). In the same section, we report exact
eigenvalues and eigenfunctions for two PDM models , $g\left( \rho \right)
=\eta \rho ^{2}$ and $g\left( \rho \right) =\eta /\rho ^{2}$. \ For the
second treatment of the radial $\rho $-dependent part (12), nevertheless, we
choose to use, in section 4, the biconfluent Heun differential form (c.f.,
e.g., \cite{35,36,37,38}). However, the implementation the biconfluent Heun
equation and its biconfluent Heun solutions necessarily means that\ these
solutions belong to the set of \emph{conditionally exact solutions} for Schr%
\"{o}dinger equation. Consequently, we report (in the same section) the 
\emph{conditionally exact }eigenvalues and eigenfunctions for two PDM
models, $g\left( \rho \right) =\lambda \rho $ and $g\left( \rho \right)
=\lambda /\rho ^{2}$. In section 5, moreover, the spectral signatures of the
eigenvalues of the one-dimensional $z$-dependent Schr\"{o}dinger part, (11)
below, on the overall spectra are reported for each of the four models used.
\ We conclude in section 5.

\section{PDM-momentum operator and minimal-coupling: cylindrical coordinates
and separability}

In this section, we start with the PDM momentum operator%
\begin{equation}
\widehat{P}\left( \overrightarrow{r}\right) =-i\left[ \overrightarrow{\nabla 
}-\frac{1}{4}\left( \frac{\overrightarrow{\nabla }m\left( \overrightarrow{r}%
\right) }{m\left( \overrightarrow{r}\right) }\right) \right] ,
\end{equation}%
suggested by Mustafa and Algadhi \cite{27} (the readers are advised to refer
to\ \cite{27} for more details on the issue) which obviously collapses into $%
\widehat{P}=-i\overrightarrow{\nabla }$ for constant mass settings. The PDM
momentum operator is to be substituted in the PDM- Schr\"{o}dinger equation

\begin{equation}
\left[ \left( \frac{\widehat{P}\left( \overrightarrow{r}\right) -e%
\overrightarrow{A}\left( \overrightarrow{r}\right) }{\sqrt{m\left( 
\overrightarrow{r}\right) }}\right) ^{2}+W\left( \overrightarrow{r}\right) %
\right] \psi \left( \overrightarrow{r}\right) =E\psi \left( \overrightarrow{r%
}\right) ;\ \ W\left( \overrightarrow{r}\right) =e\varphi \left( 
\overrightarrow{r}\right) +V\left( \overrightarrow{r}\right) ,
\end{equation}%
with electromagnetic interaction. Where $\overrightarrow{A}\left( 
\overrightarrow{r}\right) $ is the vector potential, $e\varphi \left( 
\overrightarrow{r}\right) $ is a scalar potential and $V\left( 
\overrightarrow{r}\right) $ is any other potential energy than the electric
and magnetic ones. The PDM minimal coupling\ \cite{26,27} is used in the
process. Consequently, in a straightforward manner, equation (3) would read

\begin{eqnarray}
&&\left[ -\frac{1}{m\left( \overrightarrow{r}\right) }\overrightarrow{\nabla 
}^{2}+\left( \frac{\overrightarrow{\nabla }m\left( \overrightarrow{r}\right) 
}{m\left( \overrightarrow{r}\right) ^{2}}\right) \cdot \overrightarrow{%
\nabla }+\frac{1}{4}\left( \frac{\overrightarrow{\nabla }^{2}m\left( 
\overrightarrow{r}\right) }{m\left( \overrightarrow{r}\right) ^{2}}\right) -%
\frac{7}{16}\left( \frac{\left[ \overrightarrow{\nabla }m\left( 
\overrightarrow{r}\right) \right] ^{2}}{m\left( \overrightarrow{r}\right)
^{3}}\right) +\frac{2\ i\ e}{m\left( \overrightarrow{r}\right) }%
\overrightarrow{A}\left( \overrightarrow{r}\right) \cdot \overrightarrow{%
\nabla }\right.   \notag \\
&&\quad \quad \left. +\frac{ie}{m\left( \overrightarrow{r}\right) }\left( 
\overrightarrow{\nabla \cdot }\overrightarrow{A}\left( \overrightarrow{r}%
\right) \right) -i\ e\ \overrightarrow{A}\left( \overrightarrow{r}\right)
\cdot \left( \frac{\overrightarrow{\nabla }m\left( \overrightarrow{r}\right) 
}{m\left( \overrightarrow{r}\right) ^{2}}\right) +\frac{e^{2}\overrightarrow{%
A}\left( \overrightarrow{r}\right) ^{2}}{m\left( \overrightarrow{r}\right) }%
+W\left( \overrightarrow{r}\right) \right] \psi \left( \overrightarrow{r}%
\right) =E\psi \left( \overrightarrow{r}\right) .
\end{eqnarray}%
Here, we consider the interaction of a PDM particle of charge $e$ moving in
the vector potential%
\begin{equation}
\overrightarrow{A}\left( \overrightarrow{r}\right) =\overrightarrow{A}%
_{1}\left( \overrightarrow{r}\right) +\overrightarrow{A_{2}}\left( 
\overrightarrow{r}\right) ;\left\{ 
\begin{tabular}{l}
$\overrightarrow{\nabla }\times \overrightarrow{A}_{1}\left( \overrightarrow{%
r}\right) =\overrightarrow{B}=B_{\circ }\widehat{z}\medskip $ \\ 
$\overrightarrow{\nabla }\times \overrightarrow{A}_{2}\left( \overrightarrow{%
r}\right) =0\medskip $%
\end{tabular}%
\right. ,
\end{equation}%
where a constant, uniform magnetic field $\overrightarrow{B}=B_{\circ }%
\widehat{z}$ is applied in the $z$-direction, $\overrightarrow{A}_{1}\left( 
\overrightarrow{r}\right) =\left( 0,B_{\circ }\rho /2,0\right) $ and $%
\overrightarrow{A}_{2}\left( \overrightarrow{r}\right) =\left( 0,\Phi
_{AB}/2\pi \rho ,0\right) $ are given in the cylindrical coordinates, with $%
\overrightarrow{A}_{2}\left( \overrightarrow{r}\right) $ describing the so
called Aharonov-Bohm flux field $\Phi _{AB}$ effect (c.f., e.g., \cite%
{24,31,32,33} and related references cited therein). At this point, one
should be aware that our charge $e=\pm \left\vert e\right\vert $, and we put
no restriction on its positivity or negativity as yet. Consequently, our PDM
charged particle interacts with the total vector potential

\begin{equation}
\overrightarrow{A}\left( \overrightarrow{r}\right) =\left( 0,\frac{B_{\circ }%
}{2}\rho +\frac{\Phi _{AB}}{2\pi \rho },0\right) ,
\end{equation}%
\ that satisfies the Coulomb gauge $\overrightarrow{\nabla }\cdot 
\overrightarrow{A}=0.$ Moreover, we shall use the assumptions that

\begin{equation}
m\left( \overrightarrow{r}\right) =m\left( \rho ,\varphi ,z\right)
=\,g\left( \rho \right) f\left( \varphi \right) k\left( z\right) =\,g\left(
\rho \right) ;\,f\left( \varphi \right) =k\left( z\right) =1,
\end{equation}%
and

\begin{equation}
g\left( \rho \right) W\left( \rho ,\varphi ,z\right) =V\left( \rho \right)
+V\left( \varphi \right) +V\left( z\right) =V\left( \rho \right) +V\left(
z\right) ;V\left( \varphi \right) =0,
\end{equation}%
where, $V\left( \varphi \right) =0$ assumes azimuthal symmetrization and our
PDM scalar multiplier $m\left( \overrightarrow{r}\right) =g\left( \rho
\right) $ is only radially cylindrically symmetric. This would secure
separability of the problem at hand.

Under such assumptions construction, we may now follow the conventional
textbook separation of variables and use the substitution 
\begin{equation}
\psi \left( \overrightarrow{r}\right) =\psi \left( \rho ,\varphi ,z\right)
=R\left( \rho \right) Z\left( z\right) e^{im\varphi },
\end{equation}%
(where $m=0,\pm 1,\pm 2,...,\pm \ell $ \ is the magnetic quantum number, and 
$\ell $ is angular momentum quantum number) in (4) to obtain.

\begin{gather}
\left[ \frac{R^{\prime \prime }\left( \rho \right) }{R\left( \rho \right) }%
-\left( \frac{g^{\prime }\left( \rho \right) }{g\left( \rho \right) }-\frac{1%
}{\rho }\right) \frac{R^{\prime }\left( \rho \right) }{R\left( \rho \right) }%
-\frac{1}{4}\left( \frac{g^{\prime \prime }\left( \rho \right) }{g\left(
\rho \right) }+\frac{g^{\prime }\left( \rho \right) }{\rho g\left( \rho
\right) }\right) +\frac{7}{16}\left( \frac{g^{\prime }\left( \rho \right) }{%
g\left( \rho \right) }\right) ^{2}\right.  \notag \\
\left. -\frac{m^{2}}{\rho ^{2}}-\frac{e^{2}\Phi _{AB}^{2}}{4\pi ^{2}\rho ^{2}%
}+\frac{em\Phi _{AB}}{\pi \rho ^{2}}-\frac{e^{2}B_{\circ }\Phi _{AB}}{2\pi }%
+eB_{\circ }m-\frac{e^{2}B_{\circ }^{2}\rho ^{2}}{4}+g\left( \rho \right)
E-V\left( \rho \right) \right]  \notag \\
+\left[ \frac{Z^{\prime \prime }\left( z\right) }{Z\left( z\right) }-V\left(
z\right) \right] =0.
\end{gather}%
It is obvious that this equation decouples into two parts, a purely $z$%
-dependent part

\begin{equation}
\left[ -\partial _{z}^{2}+V\left( z\right) \right] Z\left( z\right)
=k_{z}^{2}Z\left( z\right) ,
\end{equation}%
and a radial-dependent cylindrically-azimuthal part

\begin{gather}
\left[ \frac{R^{\prime \prime }\left( \rho \right) }{R\left( \rho \right) }%
-\left( \frac{g^{\prime }\left( \rho \right) }{g\left( \rho \right) }-\frac{1%
}{\rho }\right) \frac{R^{\prime }\left( \rho \right) }{R\left( \rho \right) }%
-\frac{1}{4}\left( \frac{g^{\prime \prime }\left( \rho \right) }{g\left(
\rho \right) }+\frac{g^{\prime }\left( \rho \right) }{\rho g\left( \rho
\right) }\right) +\frac{7}{16}\left( \frac{g^{\prime }\left( \rho \right) }{%
g\left( \rho \right) }\right) ^{2}\right.  \notag \\
\left. -\frac{\tilde{m}^{2}}{\rho ^{2}}+eB_{\circ }\tilde{m}-k_{z}^{2}-\frac{%
e^{2}B_{\circ }^{2}\rho ^{2}}{4}+g\left( \rho \right) E-V\left( \rho \right) %
\right] =0,
\end{gather}%
where $\alpha =\Phi _{AB}/\Phi _{\circ }$, $\Phi _{\circ }=2\pi /e$ is the
Aharonov-Bohm flux quantum (within the current units settings, of course),
and $\tilde{m}=m-\alpha $ is a new irrational magnetic quantum number that
indulges within the Aharonov-Bohm quantum number $\alpha .$ At this point,
nevertheless, one may need to get rid of the first-order derivative and
bring the radial part into the one-dimensional Schr\"{o}dinger form. In so
doing, one may use the substitution%
\begin{equation}
R\left( \rho \right) =\sqrt{\frac{g\left( \rho \right) }{\rho }}U\left( \rho
\right) ,
\end{equation}%
to obtain%
\begin{equation}
\left\{ -\frac{d^{2}}{d\rho ^{2}}+\frac{\tilde{m}^{2}-1/4}{\rho ^{2}}%
+V_{eff}\left( \rho \right) \right\} U\left( \rho \right) =\tilde{E}U\left(
\rho \right) .
\end{equation}%
Where,%
\begin{equation}
V_{eff}\left( \rho \right) =V\left( \rho \right) +\frac{e^{2}B_{\circ
}^{2}\rho ^{2}}{4}-g\left( \rho \right) E+\left[ \frac{5}{16}\left( \frac{%
g^{\prime }\left( \rho \right) }{g\left( \rho \right) }\right) ^{2}-\frac{1}{%
4}\left( \frac{g^{\prime \prime }\left( \rho \right) }{g\left( \rho \right) }%
\right) -\frac{1}{4}\left( \frac{g^{\prime }\left( \rho \right) }{\rho
\,g\left( \rho \right) }\right) \right] ,
\end{equation}%
and%
\begin{equation}
\tilde{E}=eB_{\circ }\tilde{m}-k_{z}^{2}
\end{equation}%
represents the eigenvalues of (14) which are to be used to find the
eigenvalues of the radial PDM problem at hand, i.e., to find $E_{n_{\rho
},m,\alpha }$ in (15). Obviously, result (14) retrieves the constant mass
textbook settings for $g\left( \rho \right) =1$. However, we are interested
in the case where $g\left( \rho \right) \neq const$.

On the technical mathematical side of the current methodical proposal, we
have, at our disposal, three types of Schr\"{o}dinger differential equations
to deal with.\ The $z$-dependent part of (11), the $\rho $-dependent part of
(12) and the one-dimensional $\rho $-dependent part of (14). The $\rho $%
-dependent parts (12) and (14) are to be shown useful in their own skin and
serve different PDM and/or interaction potential settings. This is clarified
in the forthcoming illustrative examples. The first batch of which consists
of exactly solvable models and the second consists of conditionally exactly
solvable models. Whereas, the $z$-dependent part of (11) will have its own
spectral signatures on the overall spectra of the decoupled problem in (3).
The strategy of our methodical proposal in handling (3) is clear, therefore.

\section{Radial cylindrical one-dimensional PDM-Schr\"{o}dinger form: a
pseudoharmonic potential and exact solvability}

In this section, consider our charged PDM particle moving in the so called
pseudoharmonic potential \cite{32}%
\begin{equation}
V\left( \rho \right) =\mathcal{V}_{_{1}}\rho ^{2}+\frac{\mathcal{V}_{_{2}}}{%
\rho ^{2}}-2\mathcal{V}_{\circ }\text{ };\text{ \ }\mathcal{V}_{_{1}}=\frac{%
\mathcal{V}_{\circ }}{\rho _{\circ }^{2}}\text{ , \ }\mathcal{V}_{_{2}}=%
\mathcal{V}_{\circ }\rho _{\circ }^{2}
\end{equation}%
in the presence of a uniform magnetic field and an Aharonov-Bohm flux field
of (6). Where, $\mathcal{V}_{\circ }$ is the chemical potential and $\rho
_{\circ }$ is the zero point of the pseudoharmonic potential. This potential
includes both a harmonic quantum dot potential $\mathcal{V}_{_{1}}\rho ^{2}$
and antidote potential $\mathcal{V}_{_{2}}\rho ^{2}$ \cite{31,32}. The
details of quantum dots and antidotes lie far beyond the scope of the
current study and can be traced from \cite{31,32}. Such a pseudoharmonic
potential is most suited for the one-dimensional PDM-Schr\"{o}dinger form
(14) and anticipated to be exactly solvable for a sample of PDM settings.
Therefore, we treat, in what follows, some special PDM settings so that
their exact solutions are inferred from some models that are known to be
exactly solvable.

\subsection{Model-I: a radial cylindrical PDM $g\left( \protect\rho \right) =%
\protect\eta \protect\rho ^{2}$}

Consider a charged particle with radial cylindrical PDM $g\left( \rho
\right) =\eta \rho ^{2}$ moving in the pseudoharmonic potential (17), under
the influence of a uniform magnetic and an Aharonov-Bohm flux fields of (6).
Then the effective potential $V_{eff}\left( \rho \right) $ of (15) would read%
\begin{equation}
V_{eff}\left( \rho \right) =\mathcal{V}_{_{1}}\rho ^{2}+\frac{\mathcal{V}%
_{_{2}}}{\rho ^{2}}-2\mathcal{V}_{\circ }\text{ }+\frac{e^{2}B_{\circ
}^{2}\rho ^{2}}{4}-\eta E\rho ^{2}+\frac{1}{4\rho ^{2}}.
\end{equation}%
Hence, equation (14) collapse into%
\begin{equation}
\left\{ -\frac{d^{2}}{d\rho ^{2}}+\frac{\tilde{\ell}^{2}-1/4}{\rho ^{2}}+%
\frac{\left( 4\mathcal{V}_{_{1}}-4\eta E+e^{2}B_{\circ }^{2}\right) }{4}\rho
^{2}\right\} U\left( \rho \right) =E_{eff}U\left( \rho \right) ,
\end{equation}%
where%
\begin{equation}
\tilde{\ell}^{2}-1/4=\tilde{m}^{2}+\mathcal{V}_{_{2}}\Longleftrightarrow
\left\vert \tilde{\ell}\right\vert =\sqrt{\left( m-\alpha \right) ^{2}+%
\mathcal{V}_{_{2}}+1/4}.
\end{equation}%
Equation (19) is, in fact, the well know two-dimensional radial cylindrical
harmonic oscillator problem (c.f., e.g., \cite{34}) that admits exact
solution in the form of%
\begin{equation}
\text{\ }E_{eff}=\sqrt{4\mathcal{V}_{_{1}}-4\eta E+e^{2}B_{\circ }^{2}}%
\left( 2n_{\rho }+\left\vert \tilde{\ell}\right\vert +1\right) =2\mathcal{V}%
_{\circ }+eB_{\circ }\left( m-\alpha \right) -k_{z}^{2}.
\end{equation}%
Which would, in turn, imply that the eigenvalues are given by%
\begin{equation}
E_{n_{\rho },m,\alpha }=\frac{1}{4\eta }\left[ 4\mathcal{V}%
_{_{1}}+e^{2}B_{\circ }^{2}-\left( \frac{2\mathcal{V}_{\circ }+eB_{\circ
}\left( m-\alpha \right) -k_{z}^{2}}{2n_{\rho }+1+\sqrt{\left( m-\alpha
\right) ^{2}+\mathcal{V}_{_{2}}+1/4}}\right) ^{2}\right] 
\end{equation}%
and radial wavefunctions are obtained in a similar manner to read

\begin{equation}
R_{n_{\rho },m,\alpha }\left( \rho \right) \sim \rho ^{1+\left\vert \tilde{%
\ell}\right\vert }\exp \left( -\frac{\sqrt{e^{2}B_{\circ }^{2}+4\mathcal{V}%
_{_{2}}-4\eta E_{n_{\rho },m,\alpha }}}{4}\rho ^{2}\right) \ _{1}F_{1}\left(
-n_{\rho };\left\vert \tilde{\ell}\right\vert +1;\frac{\sqrt{e^{2}B_{\circ
}^{2}+4\mathcal{V}_{_{2}}-4\eta E_{n_{\rho },m,\alpha }}}{2}\rho ^{2}\right) 
\end{equation}

\subsection{Model-II: a radial cylindrical PDM $g\left( \protect\rho \right)
=\protect\eta /\protect\rho ^{2}$}

A charged particle with radial cylindrical PDM $g\left( \rho \right) =\eta
/\rho ^{2}$ moving in the pseudopotential field (17), along with a uniform
magnetic and an Aharonov-Bohm flux fields of (6), would imply that equation
(14) be rewritten as%
\begin{equation}
\left\{ -\frac{d^{2}}{d\rho ^{2}}+\frac{\tilde{\ell}^{2}-1/4}{\rho ^{2}}+%
\frac{\left( 4\mathcal{V}_{_{1}}+e^{2}B_{\circ }^{2}\right) }{4}\rho
^{2}\right\} U\left( \rho \right) =E_{eff}U\left( \rho \right) ,
\end{equation}%
where%
\begin{equation}
\tilde{\ell}^{2}-1/4=\tilde{m}^{2}+\mathcal{V}_{_{2}}-\eta
E\Longleftrightarrow \left\vert \tilde{\ell}\right\vert =\sqrt{\left(
m-\alpha \right) ^{2}+\mathcal{V}_{_{2}}-\eta E+1/4}.
\end{equation}%
Equation (24) is, again, in the form of the well known two-dimensional\
radial cylindrical harmonic oscillator and admits the exact solution%
\begin{equation}
\text{\ }E_{eff}=\sqrt{4\mathcal{V}_{_{1}}+e^{2}B_{\circ }^{2}}\left(
2n_{\rho }+\sqrt{\left( m-\alpha \right) ^{2}+\mathcal{V}_{_{2}}-\eta E+1/4}%
+1\right) =2\mathcal{V}_{\circ }+eB_{\circ }\left( m-\alpha \right)
-k_{z}^{2},
\end{equation}%
to yield the eigenvalues%
\begin{equation}
E_{n_{\rho },m,\alpha }=\frac{1}{\eta }\left\{ \left( m-\alpha \right) ^{2}+%
\mathcal{V}_{_{2}}+1/4-\left[ \frac{2\mathcal{V}_{\circ }+eB_{\circ }\left(
m-\alpha \right) -k_{z}^{2}}{\sqrt{4\mathcal{V}_{_{1}}+e^{2}B_{\circ }^{2}}}%
-\left( 2n_{\rho }+1\right) \right] ^{2}\right\} 
\end{equation}%
and the corresponding radial eigenfunctions%
\begin{equation}
R_{n_{\rho },m,\alpha }\left( \rho \right) \sim \rho ^{-1+\left\vert 
\widetilde{\ell }\ \right\vert }\exp \left( -\frac{\sqrt{e^{2}B_{\circ
}^{2}+4\mathcal{V}_{_{2}}}}{4}\rho ^{2}\right) \ _{1}F_{1}\left( -n_{\rho
};\left\vert \widetilde{\ell \ }\ \right\vert +1;\frac{\sqrt{e^{2}B_{\circ
}^{2}+4\mathcal{V}_{_{2}}}}{2}\rho ^{2}\right) 
\end{equation}

\section{The biconfluent Heun-type radial cylindrical PDM-Schr\"{o}dinger
form: Killingbeck-type potential and conditionally exact solvability}

In this section, we use the radial cylindrical PDM-Schr\"{o}dinger form (12)
and consider a generalized Killingbeck potential field (e.g., \cite{40}) of
the form%
\begin{equation}
V\left( \rho \right) =V_{_{0}}+V_{_{1}}\rho +V_{_{2}}\rho ^{2}+\frac{V_{_{3}}%
}{\rho }+\frac{V_{_{4}}}{\rho ^{2}}.
\end{equation}%
When such potential field is substituted in (12), one obtains%
\begin{gather}
\left[ \frac{R^{\prime \prime }\left( \rho \right) }{R\left( \rho \right) }%
-\left( \frac{g^{\prime }\left( \rho \right) }{g\left( \rho \right) }-\frac{1%
}{\rho }\right) \frac{R^{\prime }\left( \rho \right) }{R\left( \rho \right) }%
-\frac{1}{4}\left( \frac{g^{\prime \prime }\left( \rho \right) }{g\left(
\rho \right) }+\frac{g^{\prime }\left( \rho \right) }{\rho g\left( \rho
\right) }\right) +\frac{7}{16}\left( \frac{g^{\prime }\left( \rho \right) }{%
g\left( \rho \right) }\right) ^{2}\right.   \notag \\
\left. -\frac{\beta ^{2}}{\rho ^{2}}+\tilde{k}^{2}-\gamma ^{2}\rho
^{2}+g\left( \rho \right) E-V_{_{1}}\rho -\frac{V_{_{3}}}{\rho }\right] =0,
\end{gather}%
where%
\begin{equation}
\begin{tabular}{ll}
$\beta ^{2}=\tilde{m}^{2}+V_{_{4}},$ \  & $\tilde{k}^{2}=eB_{\circ }\tilde{m}%
-k_{z}^{2}-V_{_{0}}\medskip ,$ \\ 
$\gamma ^{2}=\frac{e^{2}B_{\circ }^{2}}{4}\medskip +V_{_{2}},$ \ \ \ \ \  & $%
\alpha =\Phi _{AB}/\Phi _{\circ }$, $\Phi _{\circ }=2\pi /e.$%
\end{tabular}%
\end{equation}%
In the sample illustrative of examples below, we wish to benefit from the
known solutions of the biconfluent Heun equation using two different PDM
settings.

\subsection{Model-III: a radial cylindrical PDM $g\left( \protect\rho %
\right) =\protect\lambda \protect\rho $}

A charged PDM particle with radial cylindrical PDM $g\left( \rho \right)
=\lambda \rho $ moving in the potential field (29), under the influence of
both a uniform magnetic and an Aharonov-Bohm flux fields of (6), would be
described by the radial Schr\"{o}dinger equation (30) as%
\begin{equation}
\frac{R^{\prime \prime }\left( \rho \right) }{R\left( \rho \right) }-\frac{%
\beta ^{2}-3/16}{\rho ^{2}}-\gamma ^{2}\rho ^{2}-\frac{V_{_{3}}}{\rho }%
+\left( \lambda E-V_{_{1}}\right) \rho +\tilde{k}^{2}=0.
\end{equation}%
Which, in a straightforward manner, collapses into the standard
one-dimensional Schr\"{o}dinger form of the biconfluent Heun equation (c.f.,
e.g., \cite{35} and related references cited therein)%
\begin{equation}
R^{\prime \prime }\left( \rho \right) +\left[ \frac{1-\tilde{\alpha}^{2}}{%
4\rho ^{2}}-\frac{\tilde{\delta}}{2\rho }-\tilde{\beta}\rho -\rho ^{2}+%
\tilde{\gamma}-\frac{\tilde{\beta}^{2}}{4}\right] R\left( \rho \right) =0,
\end{equation}%
where%
\begin{equation}
\left\{ 
\begin{tabular}{lll}
$\left( 1-\tilde{\alpha}^{2}\right) /4=3/16-\beta ^{2}$, \ \medskip  & $-%
\tilde{\delta}/2=-V_{_{3}},$ \  & $-\tilde{\beta}=\lambda E-V_{_{1}}$ \\ 
$\gamma ^{2}=1=e^{2}B_{\circ }^{2}/4+V_{_{2}},$ & $\tilde{\gamma}-\tilde{%
\beta}^{2}/4=\tilde{k}^{2},$ \medskip  & 
\end{tabular}%
\right\} 
\end{equation}%
We now use the transformation recipe%
\begin{equation}
R\left( \rho \right) =\rho ^{\left( 1+\tilde{\alpha}^{2}\right) /2}\,\exp %
\left[ -\frac{\tilde{\beta}\rho +\rho ^{2}}{2}\right] \,U\left( \rho \right) 
\end{equation}%
in (33) to obtain the biconfluent Heun-type equation%
\begin{equation}
\rho \,U^{\prime \prime }\left( \rho \right) +\left[ 1+\tilde{\alpha}-\tilde{%
\beta}\rho -2\rho ^{2}\right] U^{\prime }\left( \rho \right) +\left\{ \left( 
\tilde{\gamma}-2-\tilde{\alpha}\right) \rho -\frac{1}{2}\left( \tilde{\delta}%
+\left[ 1+\tilde{\alpha}\right] \,\tilde{\beta}\right) \right\} U\left( \rho
\right) =0.
\end{equation}%
Which is known to admit solutions in the form of biconfluent Heun functions 
\begin{equation}
U\left( \rho \right) =H_{B}\left( \tilde{\alpha},\tilde{\beta},\tilde{\gamma}%
,\tilde{\delta};\rho \right) ,
\end{equation}%
where,%
\begin{equation}
\tilde{\gamma}-2-\tilde{\alpha}\text{\ }=2n_{\rho }\text{ ; \ }n_{\rho
}=0,1,2\cdots ,\medskip 
\end{equation}%
provides the essential quantization and 
\begin{gather}
\tilde{\gamma}=\frac{\tilde{\beta}^{2}}{4}+\tilde{k}^{2}=\frac{\left(
\lambda E-V_{_{1}}\right) ^{2}}{4}+eB_{\circ }\left( m-\alpha \right)
-k_{z}^{2}-V_{_{0}}\medskip , \\
\tilde{\alpha}=2\sqrt{\left( m-\alpha \right) ^{2}+V_{_{4}}+\frac{1}{16}}.
\end{gather}%
This would, in turn, imply that the eigenvalues are given as%
\begin{equation}
E_{n_{\rho },m,\alpha }=\frac{1}{\lambda }\left[ V_{_{1}}+2\left( 2\left[
n_{\rho }+1+\sqrt{\left( m-\alpha \right) ^{2}+V_{_{4}}+\frac{1}{16}}\right]
-eB_{\circ }\left( m-\alpha \right) +k_{z}^{2}+V_{_{0}}\right) \right] ^{1/2}
\end{equation}%
and the radial eigenfunctions are%
\begin{equation}
R_{n_{\rho },m,\alpha }\left( \rho \right) \sim \rho ^{\left( 1+\tilde{\alpha%
}^{2}\right) /2}\,\exp \left( -\frac{\tilde{\beta}\rho +\rho ^{2}}{2}\right)
\,H_{B}\left( \tilde{\alpha},\tilde{\beta},\tilde{\gamma},\tilde{\delta}%
;\rho \right) ,
\end{equation}%
Where $\tilde{\alpha}$ and $\tilde{\beta}$ are defined, respectively, in
(40) and(34). However, for more details on the biconfluent Heun the readers
are advised to refer to the sample of references \cite{35,36,37,38,39,40}.

\subsection{Model-IV: a radial cylindrical PDM $g\left( \protect\rho \right)
=\protect\lambda /\protect\rho ^{2}$}

For a charged PMD particle with $g\left( \rho \right) =\lambda /\rho ^{2}$
moving in the vicinity of the three fields above (i.e., the potential of
(29), the uniform magnetic and the Aharonov-Bohm flux fields of (6)), the
radial Schr\"{o}dinger equation (30) along with the substitution (13) would
collapse into%
\begin{equation}
\,\frac{U^{\prime \prime }\left( \rho \right) }{\,U\left( \rho \right) }-%
\frac{\xi ^{2}}{\rho ^{2}}-\gamma ^{2}\rho ^{2}-V_{_{1}}\rho -\frac{V_{_{3}}%
}{\rho }+\tilde{k}^{2}=0\text{ };\text{ \ }\xi ^{2}=\beta ^{2}-\lambda E.
\end{equation}%
Which, in a straight forward manner, reduces to%
\begin{equation}
U^{\prime \prime }\left( \rho \right) +\left[ \frac{1-\tilde{\alpha}^{2}}{%
4\rho ^{2}}-\frac{\tilde{\delta}}{2\rho }-\tilde{\beta}\rho -\rho ^{2}+%
\tilde{\gamma}-\frac{\tilde{\beta}^{2}}{4}\right] U\left( \rho \right) =0,
\end{equation}%
where%
\begin{equation}
\left\{ 
\begin{tabular}{ll}
$\left( 1-\tilde{\alpha}^{2}\right) /4=-\xi ^{2}=\lambda E-\left( m-\alpha
\right) ^{2}-V_{_{4}}$, \ \medskip  & $-\tilde{\delta}/2=-V_{_{3}},$ \ $%
\tilde{\beta}=V_{_{1}},$ \\ 
$\tilde{\gamma}-\tilde{\beta}^{2}/4=\tilde{k}^{2}=eB_{\circ }\left( m-\alpha
\right) -k_{z}^{2}-V_{_{0}}$ & $\gamma ^{2}=1=e^{2}B_{\circ }^{2}/4+V_{_{2}},
$ \medskip 
\end{tabular}%
\right\} 
\end{equation}%
Next, we use a transformation recipe similar to (35) and substitute%
\begin{equation}
U\left( \rho \right) =\rho ^{\left( 1+\tilde{\alpha}^{2}\right) /2}\,\exp %
\left[ -\left( \tilde{\beta}\rho +\rho ^{2}\right) /2\right] \,Y\left( \rho
\right) 
\end{equation}%
in (44) to obtain a biconfluent Heun-type equation%
\begin{equation}
\rho \,Y^{\prime \prime }\left( \rho \right) +\left[ 1+\tilde{\alpha}-\tilde{%
\beta}\rho -2\rho ^{2}\right] Y^{\prime }\left( \rho \right) +\left\{ \left( 
\tilde{\gamma}-2-\tilde{\alpha}\right) \rho -\frac{1}{2}\left( \tilde{\delta}%
+\left[ 1+\tilde{\alpha}\right] \,\tilde{\beta}\right) \right\} Y\left( \rho
\right) =0.
\end{equation}%
Which admits solutions in the form of biconfluent Heun functions 
\begin{equation}
Y\left( \rho \right) =H_{B}\left( \tilde{\alpha},\tilde{\beta},\tilde{\gamma}%
,\tilde{\delta};\rho \right) .
\end{equation}%
provided that%
\begin{equation}
\tilde{\gamma}-2-\tilde{\alpha}\text{\ }=2n_{\rho }\text{ ; \ }n_{\rho
}=0,1,2\cdots ,\medskip 
\end{equation}%
gives again the essential quantization. Where, in this case,%
\begin{gather}
\tilde{\gamma}=\frac{\tilde{\beta}^{2}}{4}+\tilde{k}^{2}=eB_{\circ }\left(
m-\alpha \right) -k_{z}^{2}-V_{_{0}}+\frac{V_{_{1}}^{2}}{4},\medskip  \\
\tilde{\alpha}=\sqrt{1+4\left[ \left( m-\alpha \right) ^{2}+V_{_{4}}-\lambda
E\right] }.
\end{gather}%
This would, in turn, imply that the eigenvalues are given by%
\begin{equation}
E_{n_{\rho },m,\alpha }=\frac{1}{\lambda }\left\{ \left( m-\alpha \right)
^{2}+V_{_{4}}+\frac{1}{4}-\left[ 2\left( n_{\rho }+1\right)
+k_{z}^{2}+V_{_{0}}-\frac{V_{_{1}}^{2}}{4}-eB_{\circ }\left( m-\alpha
\right) \right] ^{2}\right\} ,
\end{equation}%
and the radial eigenfunctions are%
\begin{equation}
R_{n_{\rho },m,\alpha }\left( \rho \right) \sim \rho ^{\left( \tilde{\alpha}%
^{2}-2\right) /2}\,\exp \left( -\frac{\tilde{\beta}\rho +\rho ^{2}}{2}%
\right) \,H_{B}\left( \tilde{\alpha},\tilde{\beta},\tilde{\gamma},\tilde{%
\delta};\rho \right) .
\end{equation}%
Where $\tilde{\alpha}$ and $\tilde{\beta}$ are defined, respectively, in
(51) and (45).

In the two examples reported above, III and IV, it is obvious that the exact
analytical solutions offered by the biconfluent Heun-type equations belong
to the set of PDM-Schr\"{o}dinger equations that are \emph{conditionally
exactly solvable}. This is mandated by the condition $\gamma
^{2}=1=e^{2}B_{\circ }^{2}/4+V_{_{2}}$ in (34) and again in (45). This
would, effectively, imply that $V_{_{2}}=1-e^{2}B_{\circ }^{2}/4$ is a
condition imposed by the exact solvability of the biconfluent Heun-type
equation that renders our radial PDM-Schr\"{o}dinger equation (12) \emph{%
conditionally exactly solvable}. Whereas, in Model-II of the preceding
section, we have used the same mass setting but not the same condition
imposed upon Model-IV above. That is why the results for the two models are
not the same as should be expected.

\section{Spectral signatures of the one-dimensional $z$-dependent Schr\"{o}%
dinger part on the overall spectra}

In this section, we shall include the $z$-dependent part (11) of the PDM Schr%
\"{o}dinger equation in (10) 
\begin{equation*}
\left[ -\partial _{z}^{2}+V\left( z\right) \right] Z\left( z\right)
=k_{z}^{2}Z\left( z\right) ,
\end{equation*}%
and explore its contribution to the spectra of the four examples discussed
above. On the mathematical theoretical side, nevertheless, we may consider
any of the conventional textbook exactly solvable one-dimensional Schr\"{o}%
dinger equations and indulge their signatures on the overall spectra.
Therefore, there exist a large number of feasible one-dimensional potentials
that may contribute to the problem at hand. However, for the sake of
clarification and illustration of the current methodical proposal,\ we only
choose two one-dimensional potentials, an infinite potential well and a
Morse-type oscillator potential.

\subsection{Case 1: Infinite potential well}

Let us assume that our charged PDM particle is also bound to move within an
impenetrable potential well of width $L$ on the $z$-axis, i.e.,%
\begin{equation}
V\left( z\right) =\left\{ 
\begin{tabular}{ll}
$0$ & ; $0<z<L$ \\ 
$\infty $ & ; elsewhere%
\end{tabular}%
\right. .
\end{equation}%
This would by, the textbook boundary conditions, manifest an exact solution
in the form of%
\begin{equation}
Z\left( z\right) \sim \sin \left( k_{z}z\right) \Longrightarrow
k_{z}L=\left( n_{z}+1\right) \pi \Longrightarrow k_{z}^{2}=\frac{\left(
n_{z}+1\right) ^{2}\pi ^{2}}{L^{2}}\text{ };\text{ }n_{z}=0,1,2,\cdots .
\end{equation}%
Under such settings, the total eigenenergies and eigenfunctions of the four
examples above are, respectively, in order.

For the two \emph{exactly solvable} models, I and II, we get%
\begin{equation}
E_{n_{\rho },m,\alpha ,n_{z}}=\frac{1}{4\eta }\left[ 4\mathcal{V}%
_{_{1}}+e^{2}B_{\circ }^{2}-\left( \frac{2\mathcal{V}_{\circ }+eB_{\circ
}\left( m-\alpha \right) -\left( n_{z}+1\right) ^{2}\pi ^{2}/L^{2}}{2n_{\rho
}+1+\sqrt{\left( m-\alpha \right) ^{2}+\mathcal{V}_{_{2}}+1/4}}\right) ^{2}%
\right] ,
\end{equation}%
\begin{eqnarray}
\psi _{n_{\rho },m,\alpha ,n_{z}}\left( \rho ,\varphi ,z\right) &=&\mathcal{N%
}\sin \left( \frac{\left( n_{z}+1\right) \pi }{L}z\right) \rho
^{1+\left\vert \tilde{\ell}\right\vert }\exp \left( -\frac{\sqrt{%
e^{2}B_{\circ }^{2}+4\mathcal{V}_{_{2}}-4\eta E_{n_{\rho },m,\alpha ,n_{z}}}%
}{4}\rho ^{2}\right) \medskip  \notag \\
&&\times \ _{1}F_{1}\left( -n_{\rho };\left\vert \tilde{\ell}\right\vert +1;%
\frac{\sqrt{e^{2}B_{\circ }^{2}+4\mathcal{V}_{_{2}}-4\eta E_{n_{\rho
},m,\alpha ,n_{z}}}}{2}\rho ^{2}\right) e^{im\varphi }.
\end{eqnarray}%
for Model-I, and

\begin{equation}
E_{n_{\rho },m,\alpha ,n_{z}}=\frac{1}{\eta }\left\{ \left( m-\alpha \right)
^{2}+\mathcal{V}_{_{2}}+1/4-\left[ \frac{2\mathcal{V}_{\circ }+eB_{\circ
}\left( m-\alpha \right) -\left( n_{z}+1\right) ^{2}\pi ^{2}/L^{2}}{\sqrt{4%
\mathcal{V}_{_{1}}+e^{2}B_{\circ }^{2}}}-\left( 2n_{\rho }+1\right) \right]
^{2}\right\} ,
\end{equation}%
\begin{eqnarray}
\psi _{n_{\rho },m,\alpha ,n_{z}}\left( \rho ,\varphi ,z\right)  &=&\mathcal{%
N}\sin \left( \frac{\left( n_{z}+1\right) \pi }{L}z\right) \rho
^{-1+\left\vert \widetilde{\ell }\ \right\vert }\exp \left( -\frac{\sqrt{%
e^{2}B_{\circ }^{2}+4\mathcal{V}_{_{2}}}}{4}\rho ^{2}\right) \medskip  
\notag \\
&&\times \ _{1}F_{1}\left( -n_{\rho };\left\vert \widetilde{\ell \ }\
\right\vert +1;\frac{\sqrt{e^{2}B_{\circ }^{2}+4\mathcal{V}_{_{2}}}}{2}\rho
^{2}\right) e^{im\varphi }.
\end{eqnarray}%
for Model-II. Moreover, for the two \emph{conditionally exactly solvable}
models, III and IV, we obtain

\begin{equation}
E_{n_{\rho },m,\alpha ,n_{z}}=\frac{1}{\lambda }\left[ V_{_{1}}+2\left( 2%
\left[ n_{\rho }+1+\sqrt{\left( m-\alpha \right) ^{2}+V_{_{4}}+\frac{1}{16}}%
\right] -eB_{\circ }\left( m-\alpha \right) +\frac{\left( n_{z}+1\right)
^{2}\pi ^{2}}{L^{2}}+V_{_{0}}\right) \right] ,
\end{equation}
\begin{equation}
\psi _{n_{\rho },m,\alpha ,n_{z}}\left( \rho ,\varphi ,z\right) =\mathcal{N}%
\sin \left( \frac{\left( n_{z}+1\right) \pi }{L}z\right) \rho ^{\left( 1+%
\tilde{\alpha}^{2}\right) /2}\,\exp \left( -\frac{\tilde{\beta}\rho +\rho
^{2}}{2}\right) \,H_{B}\left( \tilde{\alpha},\tilde{\beta},\tilde{\gamma},%
\tilde{\delta};\rho \right) e^{im\varphi }.
\end{equation}%
for Model-III, and

\begin{equation}
E_{n_{\rho },m,\alpha ,n_{z}}=\frac{1}{\lambda }\left\{ \left( m-\alpha
\right) ^{2}+V_{_{4}}+\frac{1}{4}-\left[ 2\left( n_{\rho }+1\right) +\frac{%
\left( n_{z}+1\right) ^{2}\pi ^{2}}{L^{2}}+V_{_{0}}-\frac{V_{_{1}}^{2}}{4}%
-eB_{\circ }\left( m-\alpha \right) \right] ^{2}\right\} ,
\end{equation}%
\begin{equation}
\psi _{n_{\rho },m,\alpha ,n_{z}}\left( \rho ,\varphi ,z\right) =\mathcal{N}%
\sin \left( \frac{\left( n_{z}+1\right) \pi }{L}z\right) \rho ^{\left( 
\tilde{\alpha}^{2}-2\right) /2}\,\exp \left( -\frac{\tilde{\beta}\rho +\rho
^{2}}{2}\right) \,H_{B}\left( \tilde{\alpha},\tilde{\beta},\tilde{\gamma},%
\tilde{\delta};\rho \right) e^{im\varphi }.
\end{equation}%
for Model-IV.

\subsection{Case 2: A Morse-type potential}

If our charged PDM-particle is also influenced by a Morse-type potential
(c.f., e.g., \cite{41,42} )%
\begin{equation}
V\left( z\right) =D\left[ \exp \left( -2\sigma z\right) -2\exp \left(
-\sigma z\right) \right]
\end{equation}%
in the $z$-direction, would result in the exact eigenvalues and
eigenfunctions given, respectively, as%
\begin{eqnarray}
k_{z}^{2} &=&\left( \frac{\sqrt{D}}{\sigma }-n_{z}-\frac{1}{2}\right)
^{2}\medskip , \\
Z\left( z\right) &\sim &z^{k_{z}}e^{-z/2}L_{n_{z}}^{2k_{z}}\left( z\right) ,
\end{eqnarray}%
where $L_{n_{z}}^{2k_{z}}\left( z\right) $ are the Laguerre polynomials. In
this case, the total eigenenergies and eigenfunctions of the four examples
at hand are in order.

For the two \emph{exactly solvable} models, I and II, we get%
\begin{equation}
E_{n_{\rho },m,\alpha ,n_{z}}=\frac{1}{4\eta }\left[ 4\mathcal{V}%
_{_{1}}+e^{2}B_{\circ }^{2}-\left( \frac{2\mathcal{V}_{\circ }+eB_{\circ
}\left( m-\alpha \right) -\left( \frac{\sqrt{D}}{\sigma }-n_{z}-\frac{1}{2}%
\right) ^{2}\medskip }{2n_{\rho }+1+\sqrt{\left( m-\alpha \right) ^{2}+%
\mathcal{V}_{_{2}}+1/4}}\right) ^{2}\right] ,
\end{equation}%
\begin{eqnarray}
\psi _{n_{\rho },m,\alpha ,n_{z}}\left( \rho ,\varphi ,z\right) &=&\mathcal{N%
}z^{k_{z}}e^{-z/2}L_{n_{z}}^{2k_{z}}\left( z\right) \rho ^{1+\left\vert 
\tilde{\ell}\right\vert }\exp \left( -\frac{\sqrt{e^{2}B_{\circ }^{2}+4%
\mathcal{V}_{_{2}}-4\eta E_{n_{\rho },m,\alpha ,n_{z}}}}{4}\rho ^{2}\right)
\medskip  \notag \\
&&\times \ _{1}F_{1}\left( -n_{\rho };\left\vert \tilde{\ell}\right\vert +1;%
\frac{\sqrt{e^{2}B_{\circ }^{2}+4\mathcal{V}_{_{2}}-4\eta E_{n_{\rho
},m,\alpha ,n_{z}}}}{2}\rho ^{2}\right) e^{im\varphi }.
\end{eqnarray}%
for Model-I, and

\begin{equation}
E_{n_{\rho },m,\alpha ,n_{z}}=\frac{1}{\eta }\left\{ \left( m-\alpha \right)
^{2}+\mathcal{V}_{_{2}}+1/4-\left[ \frac{2\mathcal{V}_{\circ }+eB_{\circ
}\left( m-\alpha \right) -\left( \frac{\sqrt{D}}{\sigma }-n_{z}-\frac{1}{2}%
\right) ^{2}}{\sqrt{4\mathcal{V}_{_{1}}+e^{2}B_{\circ }^{2}}}-\left(
2n_{\rho }+1\right) \right] ^{2}\right\} ,
\end{equation}%
\begin{eqnarray}
\psi _{n_{\rho },m,\alpha ,n_{z}}\left( \rho ,\varphi ,z\right)  &=&\mathcal{%
N}z^{k_{z}}e^{-z/2}L_{n_{z}}^{2k_{z}}\left( z\right) \rho ^{-1+\left\vert 
\widetilde{\ell }\ \right\vert }\exp \left( -\frac{\sqrt{e^{2}B_{\circ
}^{2}+4\mathcal{V}_{_{2}}}}{4}\rho ^{2}\right) \medskip   \notag \\
&&\times \ _{1}F_{1}\left( -n_{\rho };\left\vert \widetilde{\ell \ }\
\right\vert +1;\frac{\sqrt{e^{2}B_{\circ }^{2}+4\mathcal{V}_{_{2}}}}{2}\rho
^{2}\right) e^{im\varphi }.
\end{eqnarray}%
for Model-II. Likewise, for the two \emph{conditionally exactly solvable}
models, III and IV, we obtain

\begin{equation}
E_{n_{\rho },m,\alpha ,n_{z}}=\frac{1}{\lambda }\left[ V_{_{1}}+2\left( 2%
\left[ n_{\rho }+1+\sqrt{\left( m-\alpha \right) ^{2}+V_{_{4}}+\frac{1}{16}}%
\right] -eB_{\circ }\left( m-\alpha \right) +\left( \frac{\sqrt{D}}{\sigma }%
-n_{z}-\frac{1}{2}\right) ^{2}+V_{_{0}}\right) \right] ,
\end{equation}%
\begin{equation}
\psi _{n_{\rho },m,\alpha ,n_{z}}\left( \rho ,\varphi ,z\right) =\mathcal{N}%
z^{k_{z}}e^{-z/2}L_{n_{z}}^{2k_{z}}\left( z\right) \rho ^{\left( 1+\tilde{%
\alpha}^{2}\right) /2}\,\exp \left( -\frac{\tilde{\beta}\rho +\rho ^{2}}{2}%
\right) \,H_{B}\left( \tilde{\alpha},\tilde{\beta},\tilde{\gamma},\tilde{%
\delta};\rho \right) e^{im\varphi }.
\end{equation}%
for Model-III, and

\begin{equation}
E_{n_{\rho },m,\alpha ,n_{z}}=\frac{1}{\lambda }\left\{ \left( m-\alpha
\right) ^{2}+V_{_{4}}+\frac{1}{4}-\left[ 2\left( n_{\rho }+1\right) +\left( 
\frac{\sqrt{D}}{\sigma }-n_{z}-\frac{1}{2}\right) ^{2}+V_{_{0}}-\frac{%
V_{_{1}}^{2}}{4}-eB_{\circ }\left( m-\alpha \right) \right] ^{2}\right\} ,
\end{equation}%
\begin{equation}
\psi _{n_{\rho },m,\alpha ,n_{z}}\left( \rho ,\varphi ,z\right) =\mathcal{N}%
z^{k_{z}}e^{-z/2}L_{n_{z}}^{2k_{z}}\left( z\right) \rho ^{\left( \tilde{%
\alpha}^{2}-2\right) /2}\,\exp \left( -\frac{\tilde{\beta}\rho +\rho ^{2}}{2}%
\right) \,H_{B}\left( \tilde{\alpha},\tilde{\beta},\tilde{\gamma},\tilde{%
\delta};\rho \right) e^{im\varphi }.
\end{equation}%
for Model-IV.

\section{Concluding Remarks}

Using cylindrical coordinates $\left( \rho ,\varphi ,z\right) $, we have
considered PDM charged particles moving under not only the influence of
magnetic and Aharonov-Bohm flux fields, but also other interaction potential
fields (the reader is advised to refer to Wang \cite{43} for some
comprehensive background on the Aharonov-Bohm effect). We have implemented
the PDM-minimal-coupling\ recipe \cite{26}, along with the PDM-momentum
operator \cite{27}, and explored the separability of the problem under
radial cylindrical and azimuthal symmetrization settings. A simple
one-dimensional textbook purely $z$-dependent (11) and a purely radial $\rho 
$-dependent (12) Schr\"{o}dinger equations are obtained. For the radial $%
\rho $-dependent (12) Schr\"{o}dinger equation, we have transformed it into
a radial one-dimensional Schr\"{o}dinger form (14) and used two PDM
settings, $g\left( \rho \right) =\eta \rho ^{2}$ and $g\left( \rho \right)
=\eta /\rho ^{2}$, to report on the \emph{exact solvability} (both
eigenvalues and eigenfunctions) of our PDM charged particles moving in three
fields: the magnetic, the Aharonov-Bohm flux, and the pseudoharmonic
potential (i.e., usual settings for charged particles in quantum dots and
antidotes, e.g., \cite{31,32,33}, but here we have PDM charged particles).
This is documented in section 3. Moreover, we have used the radial $\rho $%
-dependent (12) as is and used the biconfluent Heun differential forms for
two PDM settings, $g\left( \rho \right) =\lambda \rho $ (36) and $g\left(
\rho \right) =\lambda /\rho ^{2}$ (47), to report on the \emph{conditionally
exact solvability} (both eigenvalues and eigenfunctions) of our PDM charged
particles moving in three fields: the magnetic, the Aharonov-Bohm flux, and
the generalized Killingbeck potential (reported in section 4). Yet, the
spectral signatures of the one-dimensional $z$-dependent Schr\"{o}dinger
part (11) on the overall eigenvalues and eigenfunctions reported, in section
5, using two $z$-dependent potential models (infinite potential well (54)
and Morse type potentials (64)) for each of the four examples used in
section 3 and 4. To the best of our knowledge, the current study has never
been reported elsewhere.

In the light of our experience above, our observations are in order.

We have considered two PDM-radial Schr\"{o}dinger-like equations, (12) and
(14), as a result of textbook separation of variables procedure. Our
illustrative examples were chosen in the most simplistic format so that our
approach can be clearly followed and implemented (documented in sections 3
and 4). However, the discussion above should not only be restricted to the
analytically exact (pseudoharmonic potential of (17)) or analytically
conditionally exact (generalized Killingbeck-type potential of (29) via the
biconfluent Heun equation) solvabilities reported, but also it should,
technically and/or in principle, be applicable to Schr\"{o}dinger-like
models that admit numerically exact or numerically conditionally exact
solvabilities (c.f. e.g., \cite{44} and related references cited therein).
It may very well be applied to quasi-exactly solvable models (c.f. e.g.,
Quesne \cite{45,46} and related references cited therein), or even to
non-Hermitian and pseudo-Hermitian Hamiltonian settings (c.f. e.g., \cite%
{47,48,49,50} and related references cited therein). Likewise, this would
hold true for the $z$-dependent Schr\"{o}dinger-like equation (11), where we
have only used two analytically exactly solvable models (infinite potential
well\ (54) and Morse type potentials (64)).

On the vector potential side, we have used a vector potential (5) that leads
to a uniform constant magnetic field. However, in the construction of vector
potential, the magnetic field may turn out to be position-dependent.
Therefore, it would be more appropriate and/or more general approach\ to
work with a vector potential%
\begin{equation}
\vec{A}_{1}(\vec{r})=S\left( \rho \right) \overrightarrow{\tilde{A}_{1}}(%
\vec{r})=\left( 0,\frac{B_{\circ }}{2}\rho \,S\left( \rho \right) ,0\right) ,
\end{equation}%
where $S\left( \rho \right) $ is a scalar multiplier that may absorb any
position-dependent terms that may emerge in the construction process of the
vector potential $\vec{A}_{1}(\vec{r})$ (c.f, e.g., \cite{27}). It would be
interesting to explore this problem in the near future.


\newpage 

\begin{thebibliography}{99}
\bibitem{1} O. von Roos, Phys. Rev. \textbf{B 27 }(1983) 7547.

\bibitem{2} A. de Souza Dutra, C A S Almeida, Phys Lett. \textbf{A 275}
(2000) 25.

\bibitem{3} C. Tezcan, R. Sever, \"{O}.Ye\c{s}ilta\c{s}, Int. J. Theor.
Phys, \textbf{47} (2008) 1713-1721.

\bibitem{4} O. Mustafa, S. H. Mazharimousavi, Phys. Lett. \textbf{A 358}
(2006) 259.

\bibitem{5} A. D. Alhaidari, Phys. Rev. \textbf{A 66} (2002) 042116.

\bibitem{6} O. Mustafa, Int. J. Theor. Phys. \textbf{47} (2008) 1300.

\bibitem{7} R. Bravo, M. S. Plyushchay, Phys. Rev. \textbf{D 93 }(2016)
105023.

\bibitem{8} C. Quesne, V. M. Tkachuk, J. Phys. \textbf{A} \textbf{37} (2004)
4267.

\bibitem{9} B. Bagchi, P. Gorain, C. Quesne and R. Roychoudhury, Mod. Phys.
Lett. \textbf{A 19} (2004) 2765.

\bibitem{10} Y. C. Ou, Z. Cao, Q. Shen, J. Phys. \textbf{A}: Math. Gen. 
\textbf{37} (2004) 4283-4288.

\bibitem{11} O. Mustafa, S. H. Mazharimousavi, J. Phys. \textbf{A}: Math.
Theor. \textbf{40 }(2007) 863.

\bibitem{12} R. A. C. Correa, A. de Souza Dutra, J A de Oliveira, M.G.
Garcia, J. Math. Phys. \textbf{58} (2017) 012104

\bibitem{13} O. Mustafa, S. H. Mazharimousavi, Int. J. Theor. Phys. \textbf{%
46} (2007) 1786.

\bibitem{14} A. de Souza Dutra, J. Phys. \textbf{A}: Math. Gen. \textbf{39}
(2006) 203-208.

\bibitem{15} G. T. Einevoll, P. C. Hemmer, J. Thomsen, Phys. Rev. \textbf{B} 
\textbf{42}, (1990) 3485 .

\bibitem{16} R. Koc, G. Sahinoglu, M. Koca, Eur. Phys. J.\textbf{B} \textbf{%
48} (2005) 583.

\bibitem{17} S. H. Mazharimousavi, O. Mustafa, Phys. Scr. \textbf{87} (2013)
055008.

\bibitem{18} B. Bagchi, A. Banerjee, C. Quesne, V. M. Tkachuk, J. \ Phys. 
\textbf{A}: Math. Gen. \textbf{38} (2005) 2929.

\bibitem{19} O. Mustafa, J. Phys. \textbf{A}: Math. Theor.\textbf{\ 46 }%
(2013) 368001.

\bibitem{20} O. Mustafa, J. Phys. \textbf{A}; Math. Theor. \textbf{48}
(2015) 225206.

\bibitem{21} B. G. da Costa, E. P. Borges, J. Math. Phys. \textbf{59} (2018)
042101.

\bibitem{22} O. Mustafa, J. Phys. \textbf{A}: Math. Theor. \textbf{43}
(2010) 385310.

\bibitem{23} O. Mustafa, J Phys \textbf{A}: Math. Theor. \textbf{44 (}2011%
\textbf{) }355303.

\bibitem{24} M. Eshghi, H. Mehraban, S. M. Ikhdair, Chin. Phys. \textbf{B 26}
(2017) 060302.

\bibitem{25} A . de Souza Dutra, \ J. A. de Oliveira, J. Phys. \textbf{A}:
Math. Theor. \textbf{42,} (2009) 025304.

\bibitem{26} O. Mustafa, J. Phys. \textbf{A}: Math. Theor. \textbf{52 }%
(2019) 148001.

\bibitem{27} O. Mustafa, Z. Algadhi, Eur. Phys. J. Plus \textbf{134} (2019)
228.

\bibitem{28} A. Das, J. Frenkel, S. H. Pereira, J. C. Taylor, Phys. Rev. 
\textbf{A} \textbf{70} (2004) 053408

\bibitem{29} I. Wayan Sudiarta, D. J. Wallace Geldart, Phys. Lett. \textbf{A
372} (2008) 3145-3148.

\bibitem{30} H. Asnani, R. Mahajan, P. Pathak, V. A. Singh, Pramana-J. Phys. 
\textbf{73} (2009) 573-580.

\bibitem{31} E. N. Bogachek, U. Landman, Phys. Rev. \textbf{B 52} (1995)
14067.

\bibitem{32} A. Cetin, Phys. Lett. \textbf{A 372 }(2008) 3852.

\bibitem{33} S. M. Ikhdair, M. Hamzavi, Physica \textbf{B}: Condensed Matter 
\textbf{407 }(2012) 4198.

\bibitem{34} O. Mustafa, J. Phys. Condensed Matter \textbf{5} (1993) 1327.

\bibitem{35} M. Eshghi, H. Mehraban, S. M. Ikhdair, Chin. Phys. \textbf{B 26}
(2017) 060302.

\bibitem{36} J. Rovder, Mat. Cas. \textbf{24} (1974) 15-20.

\bibitem{37} E. R .Arriola, A. Zarzo, J. S. Dehesa, J. Comput. Appl. Math. 
\textbf{37} (1991) 161-169.

\bibitem{38} A. Ronveaux,"Heun's Differential Equation", (Oxford University
Press, Oxford, 1995).

\bibitem{39} H. Karayer, D. Demirhan F. B\"{u}y\"{u}kk\i l\i \c{c}, Math.
Phys. \textbf{76} (2015).

\bibitem{40} H. Sobhani, H. Hassaabadi. W. S. Chung, Nucl. Phys. \textbf{A
973} (2018) 33.

\bibitem{41} A. Arda, R Sever, Commun. Theor. Phys. \textbf{58} (2012) 27.

\bibitem{42} G. Chen, Phys. Lett. \textbf{A 326} (2004) 55.

\bibitem{43} R. F. Wang, Fron. Phys. \textbf{10} (2015) 100305.

\bibitem{44} O. Mustafa, S. C. Chhajlany, Phys. Rev. \textbf{A 50} (1994)
2926.

\bibitem{45} C. Quesne, Acta Polytech. \textbf{58} (2018) 118.

\bibitem{46} C. Quesne, J. Math. Phys. \textbf{58} \ (2017) 052104.

\bibitem{47} O. Mustafa, S. H. Mazharimousavi, J. Phys. \textbf{A}: Math.
Theor. \textbf{41 }(2008) 244020.

\bibitem{48} O. Mustafa, S. H. Mazharimousavi, Czech.  J. Phys. \textbf{56 }%
(2006) 967.

\bibitem{49} O. Mustafa, M. Znojil, J. Phys. \textbf{A}: Math. Gen. \textbf{%
35 }(2002) 8929.

\bibitem{50} M. Znojil, F. Gemperle, O. Mustafa, J. Phys. \textbf{A}: Math.
Gen. \textbf{35 }(2002) 5781.
\end{thebibliography}
\end{document}